\documentclass[prl,twocolumn,superscriptaddress,nobibnotes,letterpaper]{revtex4-1}

\usepackage[pdftex]{graphicx}
\usepackage[pdftex]{epsfig}
\usepackage{amsmath}
\usepackage{amssymb}
\usepackage{amsfonts}
\usepackage{color}
\usepackage{wrapfig}
\usepackage{eucal}
\usepackage{hhline}
\usepackage{threeparttable}
\usepackage{supertabular}
\usepackage{multirow}
\usepackage{tabularx}
\usepackage{soul}

\usepackage{natbib}

\usepackage{array} 
\usepackage[colorlinks=true,allcolors=blue]{hyperref} 

\newcommand{\und}[1]{_\textrm{#1}}

\definecolor{cgreen}{rgb}{.1,.6,.1}
\definecolor{co}{rgb}{.1,.6,.6}
\definecolor{orange}{rgb}{.9,.4,.0}

\newcolumntype{C}[1]{>{\centering\arraybackslash}p{#1}}

\begin{document}
\pagenumbering{arabic}

\title{Microwave-to-optics conversion using a mechanical oscillator\\ in its quantum groundstate}\thanks{This work was published in \href{https://doi.org/10.1038/s41567-019-0673-7}{Nature Phys.\ \textbf{16}, 69--74 (2020)}.}

\author{Moritz Forsch}\thanks{These authors contributed equally to this work.}
\affiliation{Kavli Institute of Nanoscience, Department of Quantum Nanoscience, Delft University of Technology, 2628CJ Delft, The Netherlands}

\author{Robert Stockill}\thanks{These authors contributed equally to this work.}
\affiliation{Kavli Institute of Nanoscience, Department of Quantum Nanoscience, Delft University of Technology, 2628CJ Delft, The Netherlands}

\author{Andreas Wallucks}
\affiliation{Kavli Institute of Nanoscience, Department of Quantum Nanoscience, Delft University of Technology, 2628CJ Delft, The Netherlands}

\author{Igor Marinkovi\'{c}}
\affiliation{Kavli Institute of Nanoscience, Department of Quantum Nanoscience, Delft University of Technology, 2628CJ Delft, The Netherlands}

\author{Claus G\"artner}
\affiliation{Kavli Institute of Nanoscience, Department of Quantum Nanoscience, Delft University of Technology, 2628CJ Delft, The Netherlands}
\affiliation{Vienna Center for Quantum Science and Technology (VCQ), Faculty of Physics, University of Vienna, A-1090 Vienna, Austria}

\author{Richard A.\ Norte}
\affiliation{Kavli Institute of Nanoscience, Department of Quantum Nanoscience, Delft University of Technology, 2628CJ Delft, The Netherlands}
\affiliation{Department of Precision and Microsystems Engineering, Delft University of Technology, Mekelweg 2, 2628CD Delft, The Netherlands}

\author{Frank van Otten}
\affiliation{ Department of Applied Physics and Institute for Photonic Integration, Eindhoven University of Technology, P.O. Box 513, 5600MB Eindhoven, The Netherlands}

\author{Andrea Fiore}
\affiliation{ Department of Applied Physics and Institute for Photonic Integration, Eindhoven University of Technology, P.O. Box 513, 5600MB Eindhoven, The Netherlands}

\author{Kartik Srinivasan}
\affiliation{Center for Nanoscale Science and Technology, National Institute of Standards and Technology, Gaithersburg, MD 20899, USA}

\author{Simon Gr\"oblacher}
\email{s.groeblacher@tudelft.nl}
\affiliation{Kavli Institute of Nanoscience, Department of Quantum Nanoscience, Delft University of Technology, 2628CJ Delft, The Netherlands}


\begin{abstract}
Conversion between signals in the microwave and optical domains is of great interest both for classical telecommunication, as well as for connecting future superconducting quantum computers into a global quantum network. For quantum applications, the conversion has to be both efficient, as well as operate in a regime of minimal added classical noise. While efficient conversion has been demonstrated using mechanical transducers, they have so far all operated with a substantial thermal noise background. Here, we overcome this limitation and demonstrate coherent conversion between GHz microwave signals and the optical telecom band with a thermal background of less than one phonon. We use an integrated, on-chip electro-opto-mechanical device that couples surface acoustic waves driven by a resonant microwave signal to an optomechanical crystal featuring a 2.7~GHz mechanical mode. We initialize the mechanical mode in its quantum groundstate, which allows us to perform the transduction process with minimal added thermal noise, while maintaining an optomechanical cooperativity $>$1, so that microwave photons mapped into the mechanical resonator are effectively upconverted to the optical domain. We further verify the preservation of the coherence of the microwave signal throughout the transduction process.
\end{abstract}

\maketitle

Research into novel quantum technologies is receiving significant attention for its potential to fundamentally transform how we receive, process and transmit information. In particular, major endeavors into building quantum processors and quantum simulators are currently underway. Many leading efforts, including superconducting qubits~\cite{Kelly2015} and quantum dots~\cite{Watson2018} share quantum information through photons in the microwave regime. While this allows for an impressive degree of quantum control~\cite{Hofheinz2009}, it also limits the distance the information can realistically travel before being lost~\cite{Kurpiers2018}. At the same time, the field of optical quantum communication has already seen demonstrations over distance scales capable of providing real-world applications~\cite{Liao2017}. In particular, by transmitting information in the optical telecom band, fiber-based quantum networks over tens or even hundreds of kilometers can be envisaged~\cite{Boaron2018}. In order to connect several quantum computing nodes over large distances into a quantum internet~\cite{Kimble2008}, it is therefore vital to be able to convert quantum information from the microwave to the optical domain, and back.

Several promising approaches have been taken to realize such a microwave to optics converter, most notably by trying to either directly couple the fields inside a non-linear crystal~\cite{Witmer2016,Fan2018,Wang2018,Rueda2016}, by using rare earth ion doped crystals~\cite{OBrien2014}, magnons~\cite{Hisatomi2016} or mechanical systems as a transducer~\cite{Stannigel2010,Bochmann2013,Andrews2014,Bagci2014,vanLaer2018,MoaddelHaghighi2018,Suchoi2014}. Recent milestones include bi-directional operation~\cite{Vainsencher2016}, coherent coupling~\cite{Balram2016}, as well as efficient conversion~\cite{Higginbotham2018}, all of which make use of a mechanical oscillator as the transducer. While high conversion efficiency has been a particular success with some mechanically-mediated frequency converters, the demonstration of intrinsic noise sources compatible with conversion of a quantum state has remained an outstanding challenge. For quantum information protocols, particularly those that can tolerate optical loss, the requirement of subphoton added noise necessitates that the converter contains less than one thermal excitation~\cite{Zeuthen2016}. To this end, several experiments have recently demonstrated cooling of mechanical oscillators into the quantum groundstate of motion~\cite{Chan2011,Teufel2011b,Meenehan2015}. The low thermal occupation forms the basis for quantum control over mechanical states, with demonstrations including quantum state preparation~\cite{OConnell2010,Hong2017,Chu2017} and entanglement between multiple mechanical	degrees of freedom~\cite{Lee2011,Riedinger2018,Ockeloen-Korppi2018}. Reaching this occupation regime is complicated by the absorption of optical photons, while at the same time realizing sufficiently strong optomechanical cooperativity to suppress additional noise sources~\cite{Meenehan2014}. To date there has been no demonstration of a system with mechanically-mediated interfaces in both microwave and optical domains that operates in the quantum ground state.

\begin{figure*}
	\includegraphics[width = 0.95\textwidth]{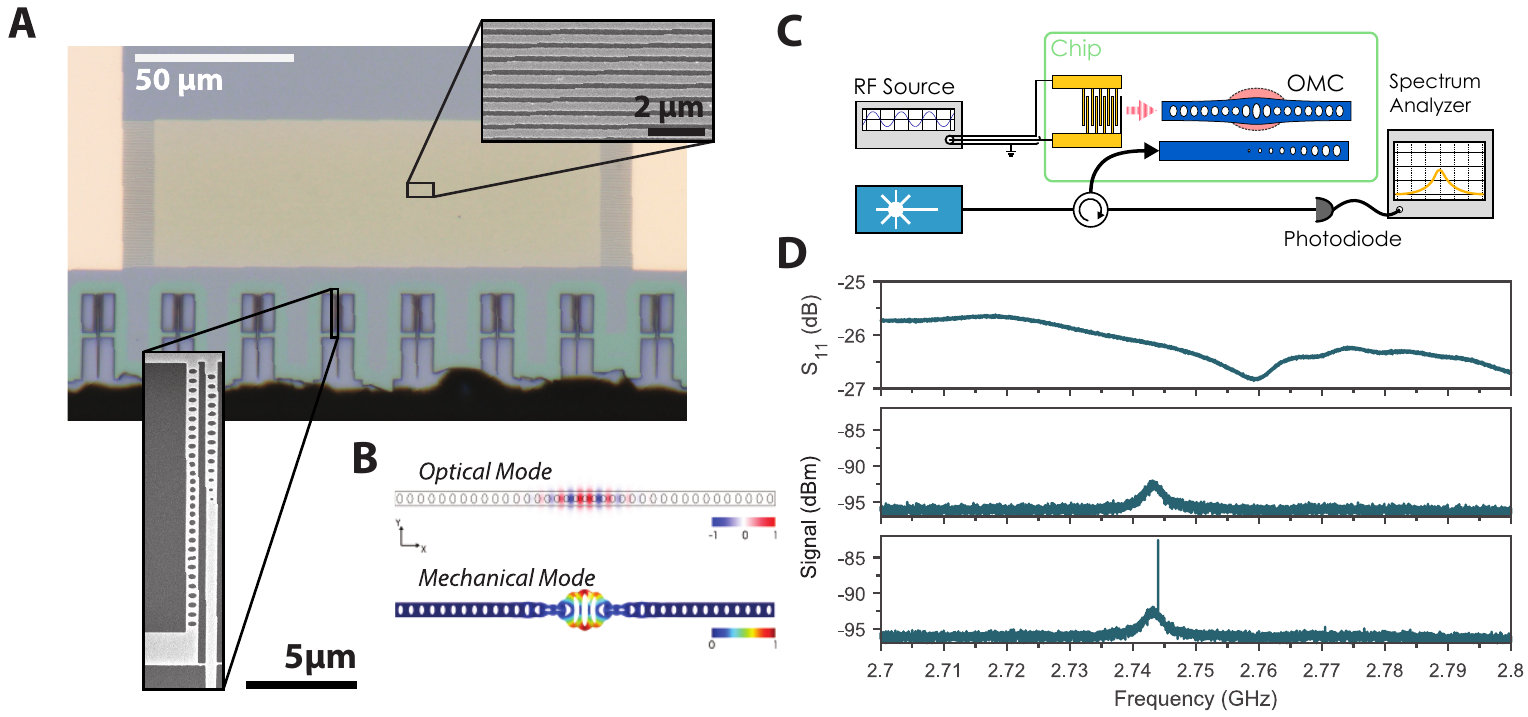}
	\caption{\textbf{Device layout and room temperature characterization.} (A) Microscope image of the transducer devices:\ Our structures are comprised of an interdigital transducer (IDT, in gold, cf. upper inset), which spans several optomechanical devices for ease of fabrication. The bottom side of the chip is directly accessible with a lensed fiber, allowing for optical access to the devices. The lower inset contains a scanning electron microscope image of an optomechanical resonator. The waveguide (right) is used for evanescently coupling light in and out of the device using the lensed fiber (accessed from the bottom, not shown). (B) Finite element simulations of the optomechanical device. The $E_y$ component of the fundamental optical mode is shown (top) alongside the displacement field of the co-localized mechanical mode oscillating around 2.7~GHz (bottom). (C) Schematic of the room temperature characterization setup. A laser is used to  address the device optically. The reflected light is then measured on a high-speed photodiode to resolve the noise spectrum around the mechanical frequency while an RF source is used to drive the IDT. (D) (Upper Panel) S$_{11}$ reflection measurement of the IDT device with a resonance at 2.76~GHz. (Lower Panels) Optical measurements of the GHz-frequency noise of the reflected light with (bottom) and without (center) the RF drive tone applied to the IDT, which results in a narrow, coherent peak in the spectrum on top of the thermal peak. The laser in these measurements is blue-detuned from cavity resonance by the mechanical frequency, $\omega\und{m}$.} \label{fig:1}
\end{figure*}

In this work, we demonstrate microwave-to-optics conversion with an electro-opto-mechanical device, which contributes less than one quantum of thermal noise. We cryogenically cool a GHz frequency piezoelectric optomechanical crystal device into its quantum groundstate of motion and excite the mechanical mode using a microwave circuit through surface acoustic waves. Crucially, our system remains in the ground state while a drive of sufficient strength for upconversion from the mechanical to the optical domain is applied. This allows for conversion of a microwave pulse to the optical telecom band in a regime where we excite on average one phonon. As our converter features a noise source containing less than one photon, optimizing the electromechanical cooperativity~\cite{OConnell2010,Moores2018}, the main efficiency bottleneck in our platform, will allow for the faithful transduction of a single photon from the microwave to optical domain.

Our microwave to optics converter consists of a one-dimensional optomechanical crystal (OMC)~\cite{Chan2012}, which is mechanically coupled to an interdigital transducer (IDT) through surface acoustic waves (see \ Fig.~\ref{fig:1}A). We fabricate the devices from a 250-nm thick GaAs layer, on a 3-$\mu$m Al$_{0.7}$Ga$_{0.3}$As sacrificial layer, both epitaxially grown on a GaAs substrate. This material combines a large refractive index ($n\und{GaAs}=3.37$ at $\lambda=1550$~nm)~\cite{Skauli2003} and a non-zero piezoelectric coefficient ($\epsilon_{14} = $-0.16~Cm$^{-2}$), with well established fabrication processes. The optomechanical device, shown in the lower inset of Fig.~\ref{fig:1}A, is designed using finite-element modeling, such that the patterned nanobeam confines light in the telecom band, while at the same time exhibiting a co-localized mechanical breathing mode at $\omega\und{m}=2\pi\times2.7$~GHz (Fig.~\ref{fig:1}B). The electro-mechanical coupling in our device is due to the piezoelectric effect that allows for the excitation of traveling acoustic waves which drive the OMC~\cite{Balram2016}. The opto-mechanial coupling, on the other hand, is facilitated by the parametric coupling between the mechanical excitation and the intracavity photon number owing to the combination of photoelastic coupling and moving boundary conditions~\cite{Chan2012}. Both effects can intrinsically operate in a bi-directional and noiseless fashion. The device fabrication consists of a two-stage lithography process to define the IDT and then pattern the nanobeams, followed by an HF etch to remove the AlGaAs sacrificial layer. A final deposition of 5~nm of AlO$\und{x}$ passivates the surfaces and reduces the effect of unwanted drive-laser absorption~\cite{Guha2017} (see Supplementary Information for details). The lower inset of Figure \ref{fig:1}A displays a scanning electron microscope image of an optomechanical device, and an evanescently coupled waveguide which provides optical-fiber access to the confined optical mode~\cite{Hong2017}.

We perform an initial characterization of the device properties at room temperature with the setup depicted in Figure \ref{fig:1}C, the results of which are presented in Figure \ref{fig:1}D. An S$_{11}$ measurement of the IDT shows a 10~MHz-wide microwave resonance centered at 2.76~GHz. The mechanical mode of the OMC is then measured by locking a laser onto the blue sideband ($\omega\und{l} = \omega\und{c}+\omega\und{m}$) of the optical cavity resonance $\omega_c$ ($\omega_c = 2\pi\times$194.3~THz, with a loaded optical quality factor $Q_\textrm{o}=3.3\times10^4$, see SI), and monitoring the high-frequency noise in the reflected signal. The peak in the noise spectrum at 2.744~GHz corresponds to the thermally-occupied mechanical mode ($\sim$1$\times 10^{3}$ phonons) and has a linewidth of several MHz. The small mismatch between the IDT and mechanical resonances of $\sim$10 MHz is a result of fabrication-based inhomogeneities. As we apply an RF tone to the IDT at the mechanical frequency $\omega\und{m}$, we observe an additional narrow peak on top of the thermal noise, corresponding to the transduced coherent signal from the IDT. The height of this peak is dependent on the RF power and the detuning from the mechanical resonance~\cite{Balram2016}.

\begin{figure}
	\includegraphics[width=1\columnwidth]{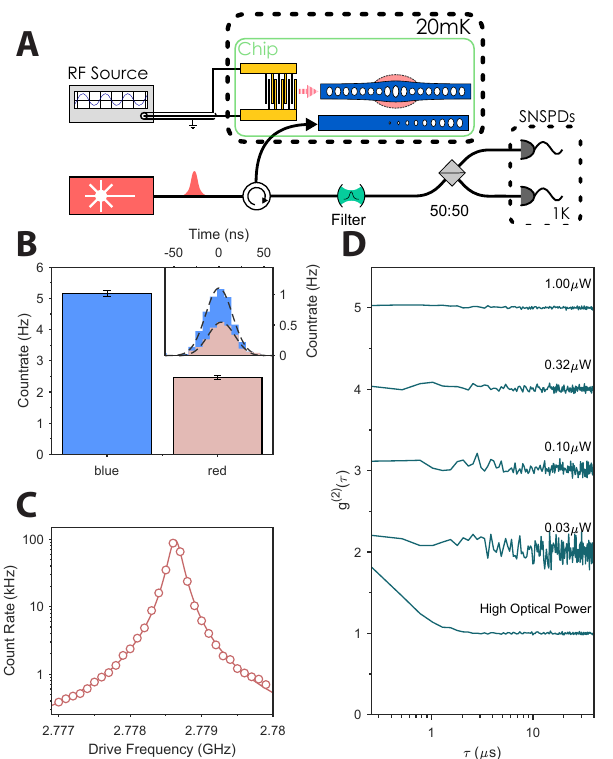}
	\caption{\textbf{Device characterization at Millikelvin temperatures}. (A) Schematic of the cryogenic experimental setup. The sample with the OMC and IDT, as well as a pair of superconducting nanowire single photon detectors (SNSPDs) are placed inside the dilution refrigerator (at 20~mK and $\sim$1~K, respectively). We lock the laser on the red sideband of our cavity and filter residual reflected pump light from the cavity, detecting photons scattered on the cavity resonance. (B) Sideband thermometry to extract the thermal occupation of the mechanical resonator. We find an occupancy $n\und{th}=0.9\pm0.01$, confirming the initialization of the device close to its quantum groundstate. The bar graph shows the integrated counts for the red and blue sideband drives as well as the corresponding histograms (inset). Errors are one standard deviation, owing to the shot noise resulting from photon counting. (C) Mechanical characterization and initial RF to telecom-band conversion at mK temperatures. We sweep the RF drive frequency with the laser locked at $\omega_c-\omega_m$ and monitor the count rate. The solid curve is a Lorentzian fit to the data, from which we extract a mechanical linewidth of 197~kHz, corresponding to a mechanical lifetime of $\sim$0.8~$\mu$s. (D) Hanbury Brown and Twiss-type measurement of the photons emitted from our cavity with 7~nW of optical input power. The second order correlations $g^{(2)}(\tau)$ are shown for a selection of the measured RF powers alongside a reference measurement with no RF drive, but high optical power (4.5~$\mu$W, bottom curve). The curves are offset for clarity. The bunching in the reference measurement results from absorption heating due to the laser drive, yielding a large thermal state of the mechanical resonator.}
	\label{fig:2}
\end{figure}

The room temperature characterization highlights that the large initial thermal occupation of the mechanical mode is a significant source of noise in this conversion process. Especially at low RF drive powers, the thermal noise dominates over the transduced signal~\cite{Balram2016}. By placing our device in a dilution refrigerator (base temperature $\sim$20~mK), we can in principle reduce the thermal occupation of the mechanical mode to $n\und{th}\approx10^{-3}$. In practice, the achievable occupation is limited by residual heating through laser absorption and finite thermalization to the cryostat~\cite{Hong2017}. With the device cooled to Millikelvin temperatures, we measure the actual thermal occupation of the mechanical mode by monitoring cavity-enhanced Stokes (blue sideband) and anti-Stokes (red sideband) scattering rates ($\Gamma\und{B}$ and $\Gamma\und{R}$, respectively), when we drive the cavity with laser pulses detuned by $\pm\omega\und{m}$~\cite{Meenehan2015}. We suppress the reflected pump light through spectral filtering, such that only scattered photons on cavity resonance are detected by superconducting nanowire single-photon detectors (SNSPD) as shown in Figure~\ref{fig:2}A. Specifically, the rates we measure are set by $\Gamma\und{B} \propto n\und{th}+1$ and $\Gamma\und{R} \propto n\und{th}$~\cite{Safavi-Naeini2012}. Figure~\ref{fig:2}B shows the histogram of the single photon count rates measured for 40~ns long pulses set to the two detunings with a peak-power of 107~nW at the device. For this power, we find an optomechanical cooperativity of $C=1.7$ (see SI)~\cite{Hill2012}. We extract a thermal occupation of $n\und{th} = 0.90\pm0.01$, which verifies the initialization of the mechanical mode close to its quantum groundstate. This value is higher than the theoretical value set by the cryostat temperature, limited by residual heating of the structure during the laser pulse~\cite{Hong2017}. A sweep of the pulse power reveals that lower occupations can be achieved (e.g.\ $n\und{th}=0.36\pm0.03$, see SI), at the cost of a lower conversion efficiency. 

We now proceed to verify the conversion from microwave to optical telecom signals at Millikelvin temperatures. Red-detuned ($\omega\und{l} = \omega\und{c}-\omega\und{m}$) optical pulses, which realize an opto-mechanical state-swap~\cite{Galland2014}, are sent into the OMC to read out the state of the mechanical mode, which is coherently excited by sweeping the frequency of an RF drive tone (1~$\mu$W) across the mechanical resonance (see Fig.~\ref{fig:2}C). The data is fitted with a Lorentzian, from which we extract a mechanical linewidth of $\gamma\und{m}/2\pi=197$~kHz, an expected improvement in mechanical quality factor of about one order of magnitude compared to the value at room temperature. Furthermore, we observe a blueshift of both the mechanical mode and the IDT resonance by about 35~MHz compared to room temperature due to the temperature dependence of the GaAs elastic constants.

While the optical absorption can be reduced through tuning of the power of the input light as well as by pulsed operation, thus far little is known about the potential heating due to the RF tone which drives the IDT. In order to investigate the amount of heating resulting from the RF drive, we measure the second order correlation function $g^{(2)}(\tau)=\langle\hat{b}^\dagger(0)\hat{b}^\dagger(\tau)\hat{b}(0)\hat{b}(\tau)\rangle/(\langle\hat{b}^\dagger(0)\hat{b}(0)\rangle\langle\hat{b}^\dagger(\tau)\hat{b}(\tau)\rangle)$ of the mechanical resonator mode~\cite{Hong2017}, by swapping mechanical excitations into optical photons at the cavity resonance frequency. Here $\hat{b}$ ($\hat{b}^\dagger$) is the annihilation (creation) operator of the mechanical mode. We expect the coherence of the RF drive to be mapped first onto the mechanical state in our resonator and then onto the light field. Any sign of heating due to the RF drive should result in bunching of the $g^{(2)}(\tau)$ during the mechanical lifetime ($\sim$1.5~$\mu$s). For this measurement, both the laser (locked onto the red sideband) and the RF source are operated in a continuous wave (CW) mode. The scattered optical photons are detected on a Hanbury Brown and Twiss (HBT)-type setup with two SNSPDs. The time $\tau$ here is the relative delay between clicks from the two detectors. We observe a near-flat $g^{(2)}(\tau)$ over the entire range of the RF power sweep (see \ Fig.~\ref{fig:2}D), clearly indicating that the coherent part of the mechanical state dominates any thermal contribution across the entire power sweep. As a reference, we also perform a measurement without an RF drive but with high enough optical power for absorption-induced heating to occur. In this case, a clear signature of photon bunching is visible, indicating, as expected, a thermal state of the mechanical resonator.

\begin{figure}
	\includegraphics[width=1\columnwidth]{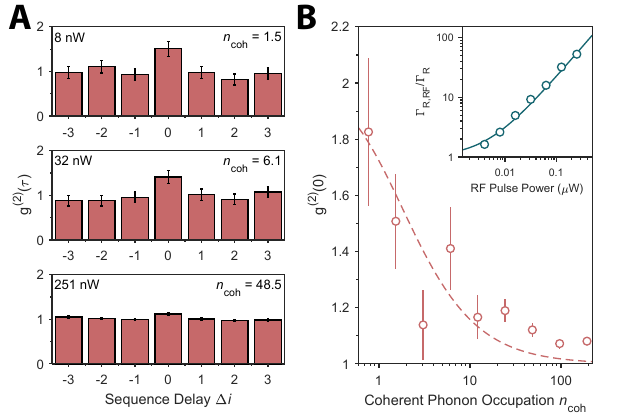}
	\caption{\textbf{Correlation measurements of the microwave-to-optical transducer in the pulsed regime.} (A) The transducer is operated such that RF drive pulses are upconverted to the optical domain using optical readout pulses. Shown are the correlations between coinciding detection events on the two single-photon detectors for photons emerging from the same $\Delta i=0$ or different $\Delta i\neq 0$ pulse sequences. The panels correspond to various coherent phonon populations. (B) The full set of $g^{(2)}(0)$ values is shown as a function of RF power applied to the IDT. The dashed curve displays the expected value of $g^{(2)}(0)$ for a displaced thermal state with the corresponding extracted coherent phonon number $n\und{coh}$ (see SI). The inset shows the relative increase in the count rate as a function of RF power with a linear fit. We use this to extract the ratio of $n\und{coh}/n\und{th}$, which allows us to demonstrate the conversion at the single coherent phonon level for the lowest powers. We can see a clear transition from a bunched (low RF power) towards a not-bunched (high RF power) second order correlation. All error bars are one standard deviation.} \label{fig:3}
\end{figure}

In order to demonstrate the potential of these devices as transducers of microwave to optical signals at the quantum level, we now operate both the RF driving and optical read-out in a pulsed mode~\cite{Riedinger2016}. We send a resonant RF pulse (1~$\mu$s long) to the IDT to excite our oscillator and access the mechanical state through a 40~ns long red-detuned state-swap pulse. This allows us to minimize the effects of heating due to optical absorption. The pulsed experiment enables us to quantify the absolute number of coherent phonons added to our initial state by comparing the scattering rate in the presence of an RF drive ($\Gamma\und{R,RF}$) to the scattering rate we obtain from the remaining thermal population ($\Gamma\und{R}$) (inset in Fig.~\ref{fig:3}B). By measuring the photon rate with and without a resonant RF drive, we recover an RF-phonon conversion effciency of $3.57\times10^{-10}$ phonons per RF photon.

Using the same HBT-type setup as above, we detect the second order correlation of the scattered photons, which allows us to compare the coincidences between detection events originating from the same ($\Delta i=0$) or different ($\Delta i \neq 0$) pulse sequences. A selection of the histograms of these correlations is shown in Figure~\ref{fig:3}A for various coherent phonon occupations ($n\und{coh}$). The full set of $g^{(2)}(0)$ values for increasing coherent phonon occupation is shown in Figure~\ref{fig:3}B. We expect the value of  $g^{(2)}(0)$ to be determined by the ratio $n\und{coh}/n\und{th}$ (cf.\ the SI). We extract this number from the relative count rate we recover with and without the RF pulse, $\Gamma\und{R,RF}/\Gamma\und{R}$, displayed in the inset of Figure~\ref{fig:3}B, making use of $\Gamma\und{R}\propto n$. The dashed curve in Figure \ref{fig:3}B displays the expected $g^{(2)}(0)$ values for our RF power sweep based on the theory for a displaced thermal state (cf.\ the SI), which are in good agreement with our measured values. An increase in RF power results in a larger coherent displacement of the thermal state, which in turn leads to a decreased value for $g^{(2)}(0)$.

\begin{figure}
	\includegraphics[width=\columnwidth]{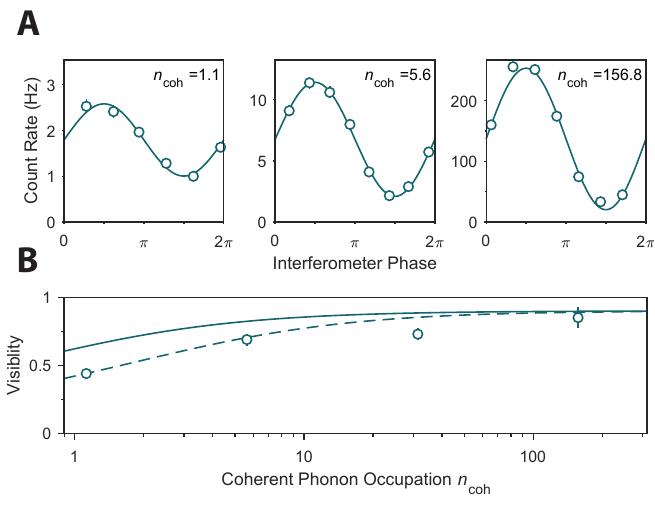}
	\caption{\textbf{Preservation of phase coherence during transduction.} (A) Interference patterns taken at different values of $n\und{coh}$. The solid curves are sinusoidal fits to the data, from which we extract the visibility. (B) Visibility as a function of the coherent phonon occupation in the mechanical resonator. The error in the visibility for the lower values of $n\und{coh}$ is smaller than the data point size. The data is overlaid with the expected visibility from the coherent and noise contributions $\sqrt{n\und{coh}/\left(n\und{coh}+n\und{noise}\right)}$. The solid line considers thermal noise only ($n\und{noise}=n\und{th}$), while the dashed line also takes any other sources into account ($n\und{noise}=n\und{th}+n\und{other}$), while both are scaled by the maximum available visibility in our setup. Error bars are one standard deviation.}
\label{fig:4}
\end{figure}

While our pulsed experiments clearly demonstrate conversion between a coherent state in the microwave and the telecom domain, they do not imply the retention of the input-state phase. In order to access the coherence of the transduction process, we use a modified version of the setup (for a detailed sketch see SI Fig.~\ref{fig:SI5}). We split the red-detuned excitation laser into two branches of a phase-stabilized Mach-Zehnder interferometer (MZI), one of which contains our device and the other an amplitude electro-optic modulator (EOM). We drive both the IDT and the EOM with a single RF source, such that a coherent transduction process results in a fixed phase relationship between the up-converted light in the two inteferometer arms. We then mix the light on a beamsplitter, matching the photon rate in the two arms. Figure~\ref{fig:4}A displays the count rate at one output port of the MZI when we vary the phase of the interferometer for several coherent phonon occupations. We observe a clear interference pattern with a visibility of 44$\pm$3\% for powers corresponding to a coherent phonon occupation of $n\und{coh}=1.1$, which increases to 85$\pm7$\% as the coherent contribution dominates over the small thermal background. Figure~\ref{fig:4}B displays these experimentally retrieved visibilities for several coherent phonon occupations, the expected modeled behavior assuming only thermal noise ($n\und{noise}=n\und{th}$, solid line), as well as additional incoherent noise sources ($n\und{noise}=n\und{th}+n\und{other}$, dashed line). These respective trends are given by $\sqrt{n\und{coh}/\left(n\und{coh}+n\und{noise}\right)}$ and scaled by the maximally achievable interference visibility in our setup of 90\%. Here, $n\und{other}$ represents the equivalent noise figure for any other source than the thermal occupation of the resonator, including imperfections in the measurement setup. We estimate the upper bound of these sources to be $n\und{other}\sim2.5$. The main contributions to this remaining part are drifts of the interferometer free spectral range over the duration of the measurement, imperfect sideband resolution ($\left(\kappa /4\omega\und{m}\right)^2=0.27$), as well as mechanical decoherence~\cite{Zeuthen2016,Andrews2014}. With this measurement, we confirm the phase-preserving nature of the conversion process down to the single phonon level.

The total efficiency of our device is the product of two parts:\ the loading efficiency of the mechanical mode from the microwave side ($3.57\times10^{-10}$, measured from the attenuator output at at the mixing chamber to the excitation of phonons in the mechanical mode) and the optical readout efficiency of the mechanical mode $1.55\times10^{-5}$. The latter one can itself be separated into two parts:\ $\eta\und{ro}=p\und{r}\times\eta\und{det}$, with $\eta\und{det}=1.41\times10^{-3}$ (see SI). The state swap probability $p\und{r}=1.1\%$ is a function of the power with which the optical readout is performed and can be increased through improvements with respect to optical absorption. Note that the current performance of our device is already sufficient to read out a non-classical state of the mechanical mode~\cite{Hong2017}. The low loading efficiency of the mechanical mode can be attributed to the design and size of the electromechanical transducer. We estimate the efficiency of transferring a SAW wave from the IDT (150~$\mu$m wide) into a single, narrow ($\sim$1~$\mu$m), suspended beam to be less than $2.5\times10^{-5}$. Additional contributions arise from the difference in the polarizations of the incoming SAW wave and the mechanical mode, the discrepancy between the IDT and mechanical frequencies, as well as the large electrical impedance of the IDT. These factors can be improved by tailoring the size and design of the electromechanical transducer specifically to the purpose of exciting the breathing mode of a single nanobeam~\cite{Wu2020}. While the small-scale piezo-resonator required to mode-match the nanobeam will necessitate careful electrical impedance matching, the required network falls into the range accessible with coplanar resonator technology~\cite{Wu2020}. The relatively small optomechanical state-swap probability and detection efficiency reported here, on the other hand, can be circumvented with post-selection techniques routinely used in quantum optics experiments~\cite{Hong2017,Marinkovic2018}. 

We have demonstrated faithful conversion of a microwave to an optical signal with only a small added thermal contribution due to the groundstate occupation of the mechanical resonator. Furthermore, these measurements show our ability to detect the displacement amplitude of the initial state in our mechanical resonator down to one phonon, (corresponding to our lowest measured RF power), marking a crucial benchmark for applications in the quantum regime~\cite{Zeuthen2016}. The device used for this experiment is a fully integrated, on-chip hybrid electro-opto-mechanical system with a mechanical mode as the transducer. We cool this mode to its quantum groundstate using a dilution refrigerator, which allows us to operate directly at the quantum noise limit. This work allows for the on-chip integration of a single-photon RF-source, such as a superconducting qubit~\cite{Gustafsson2014,Moores2018}, which paves the way for building a true quantum network over large distances, based on superconducting nodes. The device we consider here is specifically suited towards heralded entanglement generation between remote superconducting qubits, a protocol which is described in Ref.~\cite{Zeuthen2016}.

While we demonstrate quantum limited noise performance in the readout of the mechanical state, the conversion efficiency is currently limited by low microwave-phonon excitation efficiency. We would like to note that this is not a fundamental limit but rather a result of design choices for this proof-of-principle demonstration. The material itself imposes some limitations on the efficiency, such as the remaining absorption heating and the relatively low piezoelectric coupling. However, demonstrations of coherent coupling between superconducting qubits and surface acoustic waves in GaAs~\cite{Gustafsson2014,Moores2018} suggest the latter are highly suitable choices for noise-free carriers of quantum information. Importantly, our system is already operating in the range of optical drive strengths expected for mediating efficient conversion. In particular, the opto-mechanical cooperativity of $C\approx 1.7$ (see SI) is sufficient for efficiently converting GHz phonons into telecom photons.

During the submission process we became aware of related work demonstrating GaAs optomechanical crystal in the low thermal occupation regime~\cite{Ramp2019}.

\begin{acknowledgments}
We would like to thank Vikas Anant, John Davis, Mark Jenkins, and Clemens Sch\"afermeier for valuable discussions and support. We also acknowledge assistance from the Kavli Nanolab Delft, in particular from Marc Zuiddam and Charles de Boer. The sample growth was realized in the NanoLab@TU/e cleanroom facility. This project was supported by the Foundation for Fundamental Research on Matter (FOM) Projectruimte grants (15PR3210, 16PR1054), the European Research Council (ERC StG Strong-Q, 676842), and by the Netherlands Organisation for Scientific Research (NWO/OCW), as part of the Frontiers of Nanoscience program, as well as through a Vidi grant (680-47-541/994), the Gravitation program Research Center for Integrated Nanophotonics, and the ARO/LPS CQTS program.
\end{acknowledgments}

\setcounter{figure}{0}
\renewcommand{\thefigure}{S\arabic{figure}}
\setcounter{equation}{0}
\renewcommand{\theequation}{S\arabic{equation}}

\clearpage

\section{Supplementary Information}

\subsection{Device fabrication}

The samples were fabricated on an epitaxial material stack consisting of a GaAs $\langle 100\rangle$ substrate, a 3~$\mu$m thick sacrificial layer of Al$_{0.7}$Ga$_{0.3}$As, followed by a 250~nm thick device layer of GaAs. The devices are aligned with the $\langle 110 \rangle$ plane to allow propagation of the SAWs into the nanobeam. As a first step, the IDT structures, waveguides and alignment markers are defined using electron beam lithography. A stack of 5~nm Cr, 15~nm Pt, and 30~nm Au~\cite{Balram2016} is then evaporated onto the chip, with any excess removed in a lift-off process. The coplanar waveguide is impedance matched to the 50$\Omega$ coaxial cables in the setup. The OMC pattern is then aligned to the IDT structure using the markers from the previous lithography step in the second electron beam lithography step. The pattern is transferred into the device layer using a reactive ion etch process in an Alcatel GIR-300 etcher using a N$_2$/BCl$_3$/Cl$_2$ chemistry. In order to remove all organic residues, a digital etching step is performed which consists of 1~min submersion in 37\% H$_2$O$_2$, followed by thorough rinsing, 2~min submersion in 7:1 BOE, and a second round of thorough rinsing. Finally, the OMC structures are suspended by selectively removing the sacrificial layer using a 10\% HF solution, which is followed by a second round of the digital etching to remove residue from the underetch process.

\subsection{ALD passivation}

First low temperature tests of our OMCs fabricated from bare GaAs resulted in devices that were very prone to absorption of laser pulses, with varying results for $n\und{th}$ of typically many thermal phonons. We attribute the variance between different devices to a variation in the native oxide layer. Previous research~\cite{Guha2017} has indicated that a thin layer of ALD deposited aluminium oxide can reduce these absorption effects in GaAs microdiscs. We therefore proceeded to strip the native oxide off the GaAs using 7:1 BOE, followed by a 5~nm deposition of ALD AlO$_\mathrm{x}$ at 300~$^{\circ}$C. Using this method, we observe a reduction in the initial thermal occupation of our mechanical resonator from $n\und{th} \sim$ 2 -- 10 to $n\und{th}<1$.

\subsection{Optical characterization}

\begin{figure}
	\includegraphics[width=0.8\columnwidth]{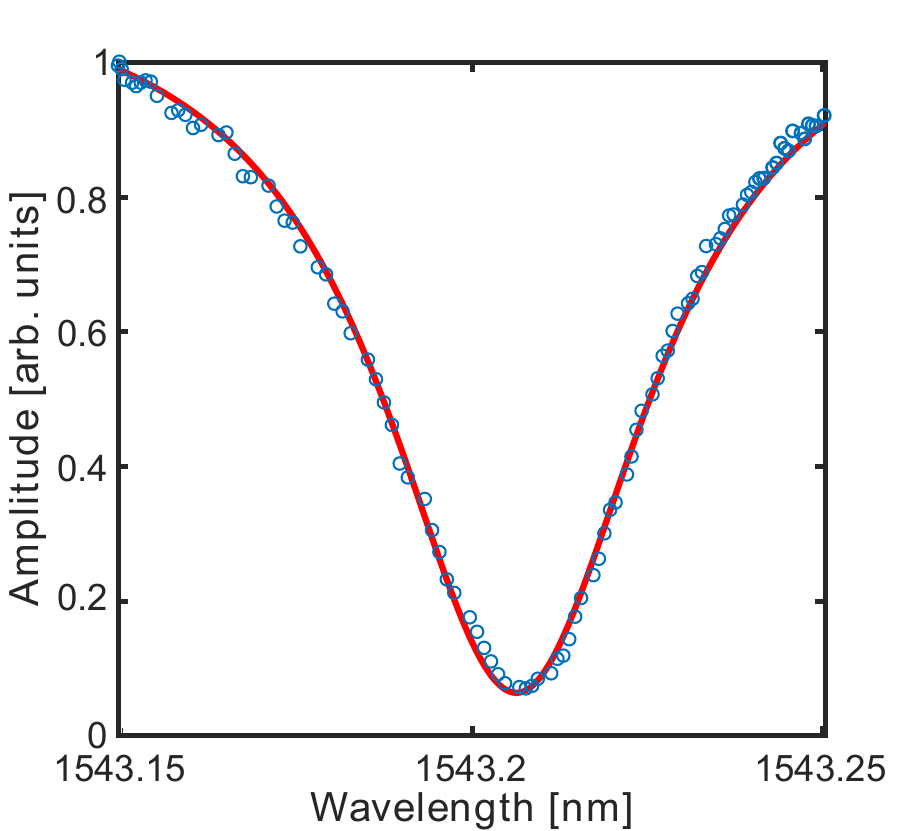}
	\caption{Scan of the optical resonance of the OMC, fitted with a Lorentzian (red solid line), which yields a linewidth of $\kappa=2\pi\times5.8$~GHz.}
	\label{fig:SI1}
\end{figure}

Optical characterization of our device is done by measuring the reflected power as we scan a tunable laser across its resonance, which is shown Fig.~\ref{fig:SI1}. The loaded linewidth is $\kappa=2\pi\times5.8$~GHz, corresponding to a quality factor of about $Q_\textrm{o}=3.3\times10^4$. Light is evanescently coupled from a central waveguide into the devices. The waveguide itself is tapered towards the edge of the chip (see Fig.~\ref{fig:1}), where we position a lensed fiber with a focal length of 14~$\mu$m. In order to calibrate the coupling efficiency from the fiber to this waveguide, we compare the off-resonant reflected power to the power reflected from a fiber mirror. This ratio gives $\eta\und{fc}^2$, resulting in $\eta\und{fc}=33\%$. In addition, we determine our optomechanical coupling rate to be $g_0/2\pi=1.3$~MHz~\cite{Hong2017}, which for 0.5~$\mu$W optical power (intra-cavity photon number $n\und{c}=280$) results in a cooperativity of $C = \frac{4g_0^2n\und{c}}{\kappa\gamma\und{m}} = 1.73$.

\begin{figure}[t]
	\includegraphics[width=\columnwidth]{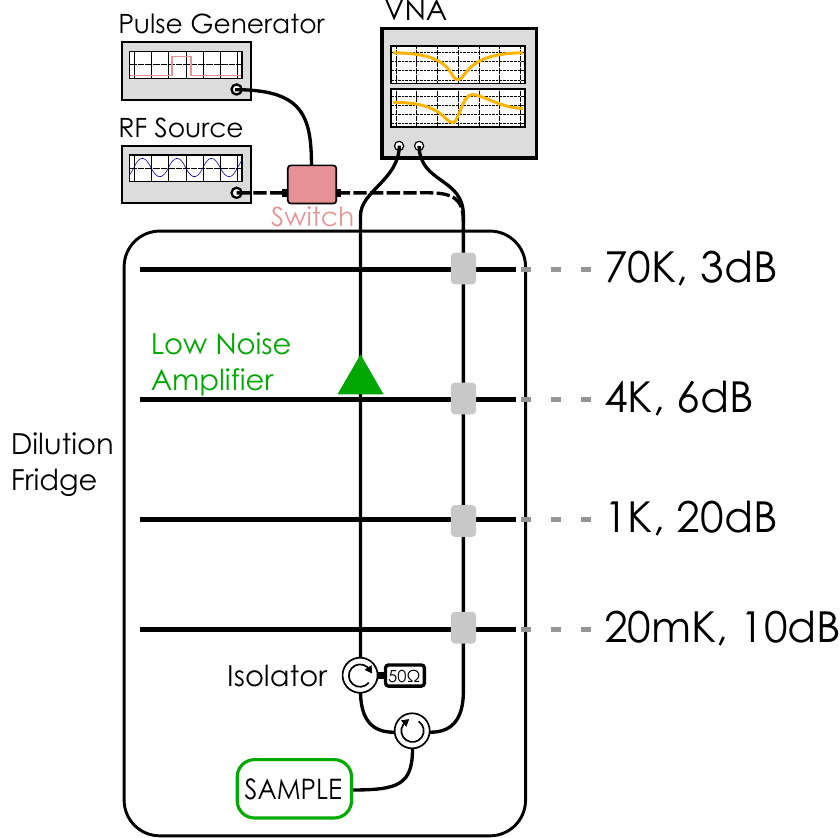}
	\caption{Schematic of the RF measurement setup. We used a vector network analyzer (VNA) to measure a reflection spectrum of our device. The signal is attenuated on the input line to reduce the noise temperature seen at the lower temperature stages and to thermalize the coaxial cable. We use a circulator to redirect the signal reflected from our device into the unattenuated output line. We use a second circulator to send any high temperature noise coming from the other stages onto a 50~$\Omega$ termination. The output signal is amplified at the 4~K stage using a low-noise amplifier. For the pulsed experiments, we used an RF source instead of the VNA and we gate the signal using an RF switch and a pulse generator.}
	\label{fig:SI2}
\end{figure}

\subsection{RF-Setup}

In order to send the microwave tone into the cryostat to our device, we attenuate the signal by several dB at the respective temperature stages in the fridge. This serves to both thermalize the coaxial cable and to reduce the noise temperature of the signal seen by the device. The specific attenuation steps are shown in Fig.~\ref{fig:SI2}. The coaxial cables connecting the stages with a temperature of $<$1~K are made from aluminium and are thus superconducting when the cryostat is cold. In order to be able to perform the S$_{11}$ measurement of our IDT device despite the strong attenuation of the input signal, we use two RF circulators at the bottom of the refrigerator. The first one connects to the device and the second one acts as an isolator sending any thermal noise coming from higher the temperature stages to the 50~$\Omega$ terminated $3^{rd}$ port of the circulator. The reflected signal from the device is then amplified using a cryogenic low-noise amplifier and sent back to the VNA. For the pulsed experiments, we use an RF source which replaces the VNA at the input of the cryostat. To generate these pulses, we send the signal through an RF switch, which is gated using a pulse generator.

\subsection{Efficiency calibration}

The calibration of the overall efficiency consists of two steps. The first part concerns the RF to phonon conversion efficiency. We send 1~$\mu$s-long RF pulses into the dilution refrigerator at a nominal power of -14~dBm, which results in, on average, 1 added coherent phonon to the resonator. Combined with the 39~dB of attenuation of the coaxial line inside the cryostat, this means we have around $2.8\times 10^9$ RF photons on the input of the device for every phonon in the resonator. This low efficiency is mostly due to the IDT not being optimized for a single OMC device, but rather being 160~$\mu$m wide, which corresponds to about 300 times the width of a single nanobeam device. Furthermore, the structure is not designed to be impedance matched to the OMC, which leads to reflections and additional losses.

We calibrate the total efficiency of our setup by sending a few-photon off-resonant laser pulse to the device. This gives the product: $\eta\und{tot}=\eta\und{fc}^2\times\eta\und{trans}\times\eta\und{QE}$, where $\eta\und{fc}$ is the fiber coupling efficiency, $\eta\und{trans}$ the transmission efficiency of the detection path (filter cavities, fiber components, etc.) , and $\eta\und{QE}$ the quantum efficiency of the detectors. We measure this total efficiency to be $8.43\times10^{-4}$. In addition we determine the coupling between the center waveguide and the nanobeam cavity, $\eta\und{dev}=\kappa\und{e}/\kappa$, to be 0.65. With the previously determined fiber-coupling efficiency of 0.33, we get a total detection efficiency $\eta\und{det} =\eta\und{fc}\times\eta\und{dev}\times\eta\und{trans}\times\eta\und{QE} = \eta\und{tot}\times\eta\und{dev}/\eta\und{fc} = 1.41\times10^{-3}$. We estimate the state readout probability (using a red-detuned pulse) to be $p\und{r}=$1.1$\%$. Within the 80~ns window considered in the correlation measurements, we find a total detector dark count rate for both detectors of 0.5~Hz.

\subsection{Sideband asymmetry}

\begin{figure}
	\includegraphics[width=1\columnwidth]{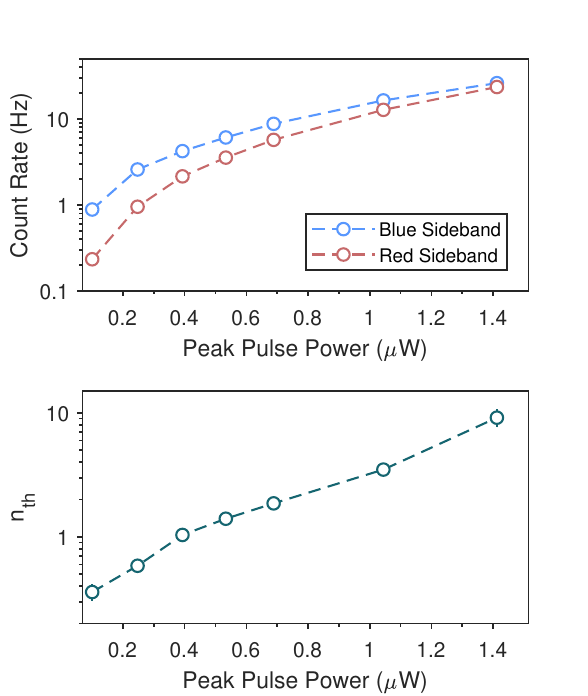}
	\caption{Sideband asymmetry for varying optical pulse powers. Upper panel:\ Count-rates for Stokes (anti-Stokes) scattering from the blue (red) cavity sideband. Lower panel:\ Extracted thermal population of the mechanical mode. Error bars showing one standard deviation are obscured by the data points. The curves in the two panels are empirical power-law fits to the data. This dataset was an early measurement of our device, recorded at slightly different settings than the measurement shown in Fig.~\ref{fig:2}b, which explains the small discrepancy in the thermal occupation.}
	\label{fig:SI3}
\end{figure}

At the 20~mK base temperature of the dilution refrigerator, we predict a thermal population of the mechanical mode of around $10^{-3}$ phonons. The occupation is increased, however by absorption of the laser pulses required to convert the microwave-frequency signal to the optical domain. Owing to this effect, we operate at an elevated occupation of $\sim$0.9 phonons for the pulsed experiments presented in the main text. Notably, the heating occurs on a rapid timescale. Absorption heating is also an important limitation for the operation of coherent opto-mechanical devices fabricated from silicon, however the slower internal dynamics compared to GaAs allow for multiple operations before the mode temperature is raised~\cite{Meenehan2015}.

In order to further illustrate the effect of absorption induced heating on our optomechanical devices, Figure~\ref{fig:SI3} displays additional sideband asymmetry measurements taken for 40~ns long Gaussian pulses with different peak powers. The upper panel displays the count rates from driving the blue and red mechanical sidebands of the optical resonance, while the lower panel displays the extracted thermal population of the mechanical resonator. Notably, for a peak power of $0.1~\mu$W, we extract a mechanical occupation of only 0.36$\pm$0.03 phonons. While operating at this power would ensure coherent transduction of even lower power microwave pulses, it also reduces the associated conversion efficiency further and increases the total measurement time beyond reasonable time scales.

\subsection{IDT - Mechanics detuning}

\begin{figure}
	\includegraphics[width=1\columnwidth]{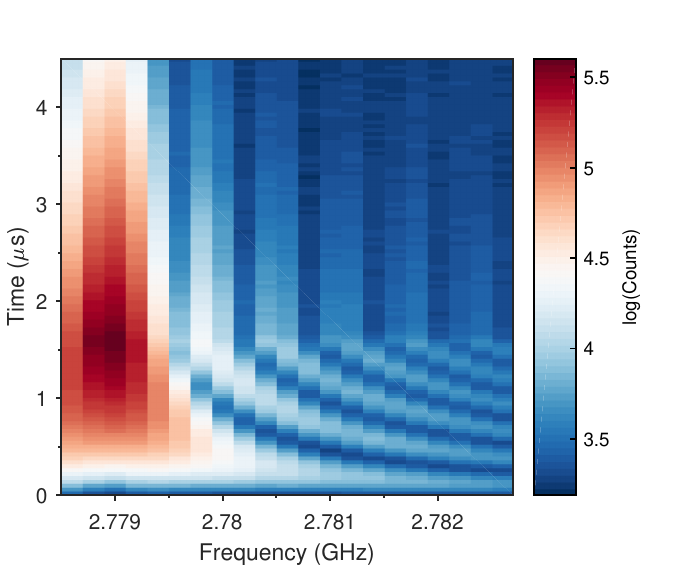}
	\caption{Time-correlated detected optical signal for continous optical driving under a detuned $1.5~\mu$s-long microwave-frequency pulse.}
	\label{fig:SI4}
\end{figure}

In Fig.~\ref{fig:2}c of the main text, we plot the mechanical resonance of the optomechanical device, which we access by monitoring the rate of anti-Stokes scattering as we vary the frequency of the coherent RF drive. This measurement can be supported by the dynamics of the mechanical mode, which we present in Fig.~\ref{fig:SI4}. For this measurement we continuously drive the red-sideband of the optical resonance at low power, and pulse the IDT with a $1.5~\mu$s-long tone, such that we continuously record the occupation of the mechanical mode. The figure displays the count-rate we record correlated to the pulse arrival time for a variety of microwave-pulse detunings from the mechanical resonance. The detuning-dependent oscillatory signal we measure during the pulse is the transient response of the mechanical oscillator to the detuned surface-acoustic wave. After the end of the pulse at $1.5~\mu$s, the mechanical resonance continuously decays to the thermal background level.

\subsection{Displaced thermal states}

In the main text we have used a displaced thermal state model to fit our data (cf.\ Fig.~\ref{fig:3}). Here we briefly derive the second order correlation function for such a state. We start with an initial thermal occupancy $n\und{th}$ of our mechanics and displace it coherently with an amplitude $\sqrt{n\und{coh}}$. Thus, our resulting mode occupation is given by  
\begin{equation}
\left\langle \hat{n} \right\rangle = n\und{coh} + n\und{th}. \label{eq:1}
\end{equation}
To derive the expected $g^{(2)}(0)$ values for a given ratio of $n\und{coh}/n\und{th}$, we follow~\cite{Vogt1991} and start with the general expression for $g^{(2)}(0)$:
\begin{equation}
g^{(2)}(0) = \frac{\left\langle \hat{a}^\dagger \hat{a}^\dagger \hat{a} \hat{a}\right\rangle}{\left\langle \hat{a}^\dagger \hat{a}\right\rangle^2} = \frac{\left\langle \hat{a}^\dagger \hat{a} \hat{a}^\dagger \hat{a} \right\rangle -\left\langle \hat{a}^\dagger \hat{a} \right\rangle}{\left\langle \hat{a}^\dagger \hat{a}\right\rangle^2}, \label{eq:2}
\end{equation}
where $\hat{a}$ and $\hat{a}^\dagger$ are annihilation and creation operators,respectively, and we used the commutation relation $[\hat{a},\hat{a}^\dagger] = 1$. Substituting $\hat{a}^\dagger \hat{a}=n$ and $\sigma^2 = \left\langle\hat{n}^2 \right\rangle -\left\langle \hat{n} \right\rangle^2$, we can simplify this to
\begin{equation}
g^{(2)}(0) = \frac{\left\langle \hat{n}^2 \right\rangle - \left\langle \hat{n} \right\rangle}{\left\langle \hat{n} \right\rangle^2} = 1+\frac{\sigma^2-\left\langle \hat{n} \right\rangle}{\left\langle \hat{n} \right\rangle^2}. \label{eq:3}
\end{equation}
For a purely thermal state, this results in
\begin{equation}
g^{(2)}\und{th}(0) = 2 = 1+\frac{\sigma^2\und{th}-n\und{th}}{n\und{th}^2}\rightarrow \sigma\und{th}^2 = n\und{th}^2 + n\und{th}. \label{eq:4}
\end{equation}
Using $\sigma^2 = \sigma\und{th}^2 +n\und{coh}(1+n\und{th})$~\cite{Vogt1991} and substituting it in Eq.~\eqref{eq:3}, we get
\begin{equation}
g^{(2)}(0) = 1+\frac{\sigma\und{th}^2+n\und{coh}(1+2n\und{th})-\left\langle \hat{n} \right\rangle}{\left\langle \hat{n} \right\rangle^2}. \label{eq:5}
\end{equation}
Substituting $\sigma\und{th}^2$ from Eq.~\eqref{eq:4}, $\left\langle \hat{n}\right\rangle$ from Eq.~\eqref{eq:1} and simplifying yields
\begin{equation}
g^{(2)}(0) = 2 - \frac{n\und{coh}^2}{\left(n\und{th} + n\und{coh}\right)^2} = 2 - \frac{1}{\left(\frac{n\und{th}}{n\und{coh}}+1\right)^2}.
\end{equation}
This result is also consistent with the one obtained by Marian et al.~\cite{Marian1993}. We directly extract the ratio $\frac{n\und{coh}}{n\und{th}}$ from $\Gamma\und{R,RF}/\Gamma\und{R} = \left(n\und{coh} + n\und{th}\right)/n\und{th}$, where $\Gamma\und{R}$ is the count rate resulting from a red-detuned pulse alone and $\Gamma\und{R,RF}$ is the count rate resulting from a red-detuned pulse, which is overlapped with the RF drive pulse, displayed in the inset of Figure~\ref{fig:3}b in the main text.

\subsection{Phase Sensitive Detection Setup}

\begin{figure}
	\includegraphics[width=\columnwidth]{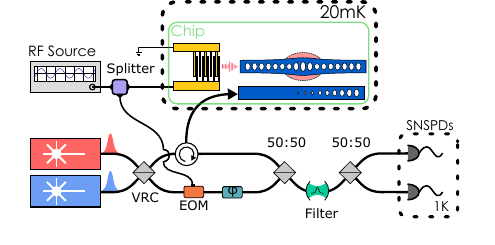}
	\caption{Phase sensitive measurement setup. We lock one laser (red) on the red sideband and the other (blue) far off resonance and split them into the two interferometer arms using a variable coupler (VRC). One RF source is used to drive both the device and the electro-optical modulator (EOM). The EOM creates sidebands on the red-detuned laser pulse, while at the same time providing a well defined phase with respect to the IDT RF drive tone. The off-resonant laser is only used to stabilize the interferometer.}
	\label{fig:SI5}
\end{figure}

In order to confirm the coherence of the transduction process from our RF source all the way to the telecom-band photons, we use an interferometer with our device in one arm and an electro-optic modulator (EOM) in the other (cf.\ Fig.~\ref{fig:SI5}). By driving both the IDT and the EOM with the same RF source, we generate optical sidebands of the drives in both interferometer arms with a well defined phase. This allows us to verify the coherence of the transduction process by measuring the interference between the two signals. We use a strong second, far detuned laser pulse (blue in Fig.~\ref{fig:SI5}) to stabilize the interferometer. This locking-drive is reflected from the photonic crystal mirror in the waveguide without resulting in intracavity photons and is delayed from the red measurement tone. The stabilization is done by measuring the transmission of this strong laser pulse at both outputs of the interferometer and feeding back the signal to a fiber stretcher ($\phi$ in the figure). We change the interferometer phase for the red (and the upconverted) light by sweeping the frequency of the locking laser over the free spectral range (approx.\ 20 GHz) of the interferometer. Both lasers, as well as other sidebands which are not resonant with the cavity, are filtered out before the SNSPDs.

\end{document}